\documentclass{aa}

\usepackage{astron}
\usepackage{graphics}
\usepackage{epsfig}

\thesaurus{11.04.1  }

\begin {document}

\titlerunning{VLT Deep I--band SBF of IC 4296}

\title{VLT Deep I--band Surface Brightness Fluctuations of IC~4296
\thanks{Based on observations performed at the European Southern Observatory, Paranal, Chile; ESO program N$^o$ 63.O-0450.}}

\author{S.Mei  \inst{1},  D. Silva \inst{2}, P.J. Quinn \inst{2}}

\institute{ 
 Observatoire Midi--Pyr\'en\'ees, 14 ave. E. Belin, 31400 Toulouse, France; mei@ast.obs-mip.fr
\and
European Southern Observatory, Karl-Schwarzschild-Strasse 2, 85748 Garching, Germany 
}

\date{Received 17 April 2000 / Accepted 5 July 2000 }

\maketitle

\begin{abstract}

\noindent From theoretical predictions, the introduction of 8--m class telescopes permits one to extend Surface Brightness Fluctuations measurements from the ground to $\approx$7000 km/s, with a precision that is comparable to current space--based measurements.
We have measured I--band SBF in IC 4296 in Abell 3565 $cz \sim 3630$ km/s with the ESO Very Large Telescope. Adopting the Tonry et al. 2000 calibration for I--band SBF we determined a distance modulus of $(\overline I_{o,k} - \overline M_{I}) = 33.44 \pm 0.17$ corresponding to a galaxy distance of 49 $\pm$ 4 Mpc. This result is consistent with the HST observation from Lauer et al. 1998: $(\overline I_{F814} - \overline M_{F814}) = 33.47 \pm 0.13$.

 \keywords{
                Galaxies: distances --
                individual: IC 4296
               }

\end{abstract}

\section{Introduction}

\noindent I--band Surface Brightness Fluctuations (SBF) have been successfully used  to measure early-type galaxy distances up to 4000 km/s from the ground and up to 7000 km/s with the Hubble Space Telescope  \cite{sod95,sod96,aj97,ton97,tho97,lau98,pah99,bla99a,ton00}. 
This method was introduced by Tonry \& Schneider 1988 (a recent review has been given by Blakeslee et al. 1999a) and is based on a simple concept. The Poissonian distribution of unresolved stars in a galaxy produces fluctuations in each pixel of the galaxy image.
The variance of these fluctuations is inversely proportional to the square of the galaxy distance. The SBF amplitude is defined like this variance normalized to the mean flux of the galaxy in each pixel \cite{ts88}. The stars that contribute the most to the SBF are then the brightest stars in the intrinsic luminosity function, typically red giants in old stellar populations.  The absolute
magnitude of the fluctuation is not a constant, but depends on the age
and metallicity of the stellar population. As such it is a strong
function of not only the photometric band in which the observations
are carried out, but also of the color of the galaxy under study.
Tonry et al. (1997) and Tonry et al. (2000) have quantified these
dependencies using an extensive sample of I and V band observations,
which are used to empirically calibrate the dependence of the I-band
fluctuation magnitude on (V-I) color. The absolute calibration is
then obtained using a set of 5 galaxies with independent Cepheid
distance. The I--band SBF absolute magnitude calibration can be written as:
\begin{equation}
\overline M_I=-1.74+4.5\ [(V-I)_0-1.15]. \label{eq:Ical}
\end{equation}
\noindent Ferrarese et al. (2000a) have obtained their own calibration based
on accurate Cepheid distances to six spiral galaxies with SBF
measurements within 1200 km/s. The Cepheid distances were part of a larger dataset of 18 distances to galaxies within the Fornax and Virgo clusters observed
by the HST Key Project on the Extragalactic Distance Scale. The two calibrations are consistent within the errors. 

\small{

\begin{table*}[!htf]

\begin{flushleft}
\begin{tabular} {|c|c|c|c|c|c|c|c|c|} \hline 
Name&RA &DEC &l&b&Type&$m_V$&$V-I$&$V_{LG} $ \\ \hline
&(2000)&(2000)&deg&deg&&mag&mag&km/s \\ \hline
IC 4296&13 36 39.46 & -33 57 59.8 &313.54 &+27.97&E&10.57&1.24&3630 \\ \hline

\end{tabular}

\end{flushleft}

\caption{General properties of IC 4296, from the web archive Simbad (http://simbad.u-strasbg.fr), and Lauer et al. 1998} \label{tab-vlt}
\end{table*}
}
\normalsize
\noindent The potential of these measurements extends to the measurements of cosmological parameters and the study of the dynamics of the universe. SBF distances have been used in measuring the Hubble constant in the HST Key project on the Extragalactic Distance Scale (Ferrarese et al. 2000a, and references therein).
Tonry and his collaborators \cite{ton00} have derived peculiar velocities from their ground based sample up to 4000~km/s and calculated bulk flows in this region. By comparing these peculiar velocities with infrared and optical
surveys, Blakeslee et al. 1999a have been able to constrain the value
of the Hubble constant and the parameter density (luminous plus dark
matter) $\beta=\Omega^{0.6}/b$, where $b$ is the linear bias \cite{bla99b}.

\noindent However, the current I--band sample is limited to 4000~km/s. The introduction of large telescopes, such as the Very Large Telescope (VLT), have opened new opportunities to extend the current sample from the ground. With the optical instrument FORS1 and FORS2 on VLT, from a theoretical point of view,  it will be possible to extend I--band SBF with high signal--to--noise and good understanding of external source contribution up to $\approx$ 7000~km/s \cite{mei00}. This implies that ground--based  measurements can be used to probe the same volume as HST measurements.

\noindent To illustrate the potential of I--band SBF measurements with the VLT, we have observed IC 4296 in Abell 3565.  This galaxy  is part of the brightest ellipticals in the Abell clusters observed in the Lauer \& Postman 1992 sample. An I--band SBF distance measurement of this galaxy, based on HST observations, has already been reported by Lauer et al. (1998). 
We compare our SBF measurements with Lauer et al. (1998) and obtain similar results and measurement precision, being our errors and external source contribution estimation comparable with HST capabilities.

\section{Observations}

IC 4296 was observed in service mode at the Very Large Telescope unit UT1 (Antu) at the European Southern Observatory in Paranal, Chile. The properties of this galaxy are summarized in Table~\ref{tab-vlt}.

\noindent I--band and V--band data were obtained, to have an accurate V--I color  to input in Tonry et al. (2000) calibration of I--band SBF fluctuations.
The nights of observation were on 19 June and  24 July 1999 for the I--band and on 1 September 1999 for the V--band.
The imaging camera was FORS1 (FOcal Reducer and low dispersion Spectrograph) with a 2048 x 2048 pixel Tektronix  CCD.

\noindent The low gain, standard resolution mode, was used, with a pixel scale 0.2$\arcsec$/pixel and field of view 6.8$\arcmin$ x 6.8$\arcmin$. 
The detector had four ports with different read out noise and gain. We show the gains for the different ports in Table~\ref{tab-fors}. The dark current is negligible ($<2e^-$/pixel/hour at nominal operating temperature).

\noindent The filter was a I--band Bessel filter. The seeing was $\approx 0.7\arcsec$ FWHM on both 19 June and 24 July.
The photometric zero-point and the extinction coefficient were extracted from  observations of Landolt standards in the I--band Kron--Cousins \cite{lan92}. In the I--band we measure a photometric zero-point $m_1=25.48 \pm 0.03$ mag  and an extinction coefficient of $0.07 \pm 0.015$ mag. 
The sky brightness was on average $m^I_{sky}=19.09$ mag/arcsec$^2$ on 19 June, $m^I_{sky}=18.26$ mag/arcsec$^2$ on 24 July.

\noindent The raw data were galaxy exposures each of 45 seconds for a total exposure time in the I--band of 7785 seconds. They were dithered by $\approx 5\arcsec$. Individual bias subtracted and flatfield corrected frames were produced and delivered by VLT Science Operations using the VLT
Data Pipeline \cite{han00}.   These individual frames were then combined. Bad pixels and cosmic rays were eliminated by a sigma clipping algorithm while combining the images by the IRAF task IMCOMBINE. Sub-pixel registration was not used to avoid the introduction of artificial statistics in the images.
The galaxy center was in the center of the CCD.
Sky estimates were done on the four corners of the CCD at $\approx 3 \arcmin$ from the galaxy center.


\begin{figure*}

\centerline{\psfig{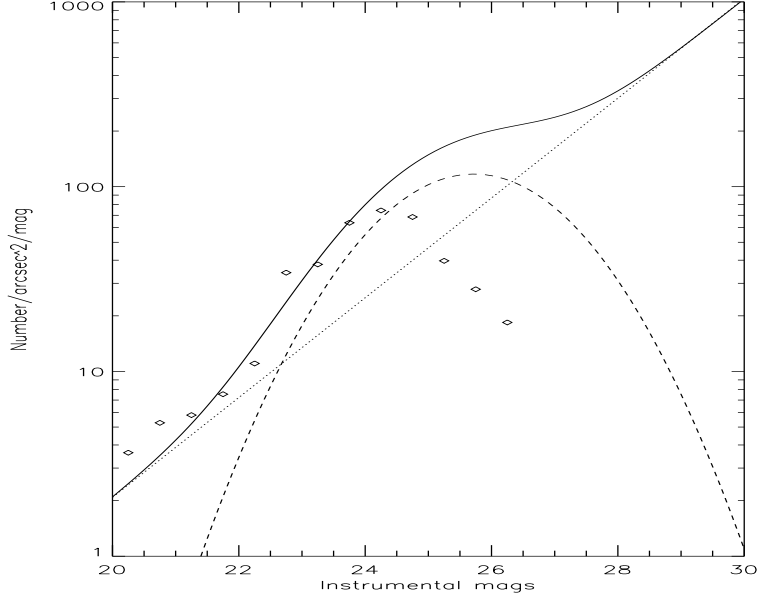}}

\caption{We show IC 4296 external source luminosity function between 34$\arcsec$ and 64$\arcsec$. The continuous line shows the sum of the globular cluster luminosity function  plus the background galaxy luminosity function. The dashed line shows the fitted globular cluster luminosity function, the dotted line the background galaxy luminosity function.} \label{fig-4296gc}

\end{figure*}

\section[Surface Brightness Fluctuations Analysis]{Surface Brightness Fluctuations Analysis}

\begin{table*}

\begin{flushleft}
\begin{tabular} {|c|c|c|} \hline 
Port&FORS1 Ron ($e^{-}$)&Gain ($e^{-}$/ADU)\\ \hline

A&5.75$\pm$0.20&2.90$\pm$0.09 \\
B&6.30$\pm$0.18&3.50$\pm$0.09 \\
C&5.93$\pm$0.18&3.08$\pm$0.09 \\
D&5.75$\pm$0.19&3.22$\pm$0.10 \\ \hline

\end{tabular}

\end{flushleft}

\caption{Read out Noise and Gain for the four ports of FORS1 on VLT UT1} \label{tab-fors}
\end{table*}


\subsection[Measure of SBF magnitudes]{Measure of SBF magnitudes}

The data were analyzed by the standard SBF extraction technique used e.g. by Tonry \& Schneider 1988.
To measure the fluctuations, a smooth galaxy profile was subtracted from to the image. It was derived by fitting galaxy isophotes to the original image, once visible external sources were subtracted.
On the residual image, additional point sources were identified, using the software tool Sextractor \cite{ber96} and subtracted.  To account for residual sky-subtraction errors it was then smoothed on a scale ten times the width of the PSF and subtracted from the image.
To make the SBF fluctuations constant across the image, the image was divided by the square root of the galaxy model.
The resulting image was divided into different annuli, in each of which the power spectrum was calculated.
The external point sources brigther than $m_{cut} = 24.7$ mag  were masked out. 
A point spread function (PSF) profile was determined from the bright stars in the image and normalized to 1 ADUs$^{-1}$. The PSF power spectrum was then calculated.
In each annulus the power spectrum of the image was calculated and normalized  to the number of non-zero points in the annulus.
The power spectrum was azimuthally averaged.

\noindent Two components contribute to the total image power spectrum: the constant power spectrum due to the white noise, $P_1$, and the power spectrum of the fluctuations and point sources that are both convolved by the PSF in the spatial domain. In the Fourier domain these latter terms are given by a constant, $P_0$, multiplied by the power spectrum of the PSF:
\begin{equation}
E_{gal}= P_0 \  E_{PSF} +P_1 .
\end{equation}


\begin{figure*}[!htf]

\centerline{\psfig{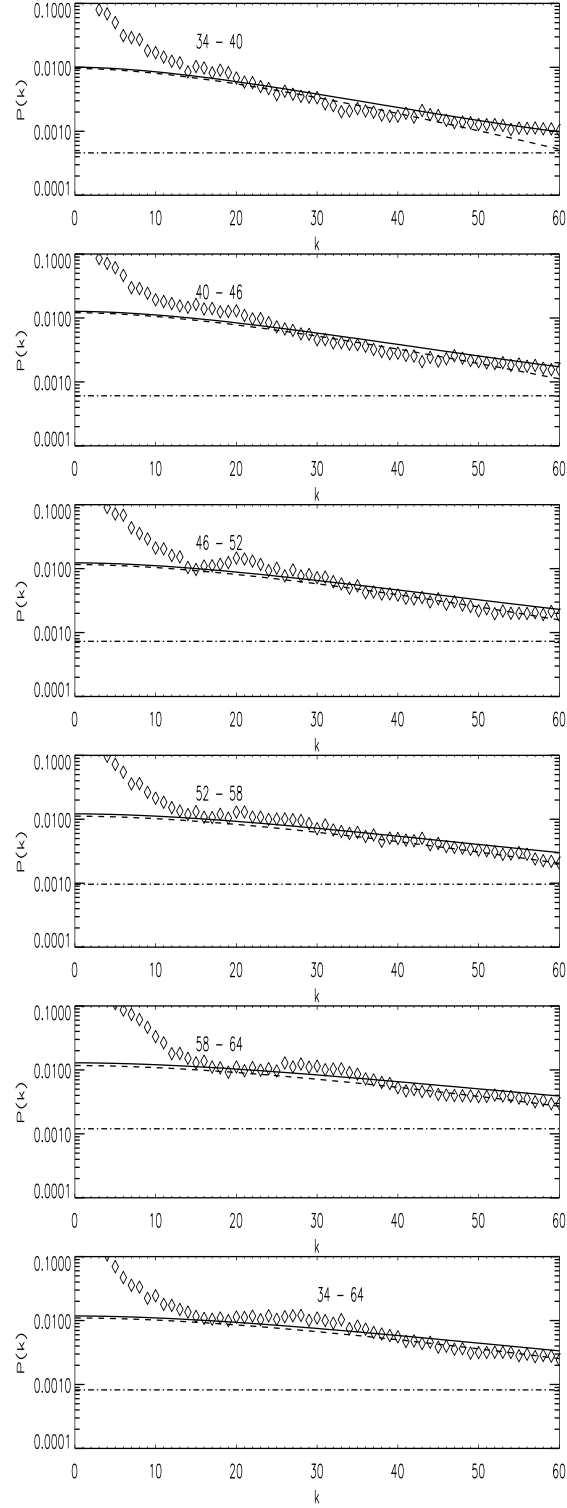}}

\caption{IC 4296 data. We show IC 4296 power spectrum as was fitted in five different annuli of width $\approx$ 6 $\arcsec$ in the galaxy and on the full field up to 64$\arcsec$. The fit of the power spectrum is given by the continue line, the PSF power spectrum by the dashed line and the dashed-dotted line is the fitted constant white noise spectrum. } \label{fig-4296spe}

\end{figure*}

\noindent The external source luminosity function for IC 4296  is shown in Fig.~\ref{fig-4296gc}.

\noindent To compute $P_0$ and $P_1$, a robust linear least squares fit was made by minimizing absolute deviation (Numerical Recipes, Press et al. 1992).
Low wave number points were excluded from the fit. In fact, they are contaminated by the galaxy subtraction and subsequent smoothing, as well as a residual sky variance contribution, as pointed out, i.e., by  Jensen et al. (1999).
The point source contribution  $P_{es}$ was estimated from the equations:
\begin{equation}
P_{es}=\sigma^2_{gc}+\sigma^2_{bg}
\end{equation}
following, i.e.,  Blakeslee \& Tonry (1995); $\sigma^2_{gc}$ is the contribution to the fluctuations given by globular clusters, $\sigma^2_{bg}$ is the contribution by background galaxies.
We adopted the following luminosity function for the globular clusters: 
\begin{equation}
N_{gc}(m) = \frac{N_{ogc}}{\sqrt{2\pi}\sigma} 
e^{\frac{-(m-m^{gc}_{peak})^2}{2\sigma^2}},
\end{equation}
\noindent with  $\sigma=1.35$. We have fitted to our detected globular cluster number counts an initial maximum likelihood  $m^{gc}_{peakI}=25.5$ and have assumed an initial galaxy distance modulus 33.47, from Lauer et al. (1998). 
From these values, we have an initial $M^{gc}_{peakI}=-7.97$. This value is consistent with the $M^{gc}_{peakI} = -7.75$ measured  by Jensen et al. (1998) in the $I_{F814W}$ filter ($I_{KC} - I_{F814W} \approx -0.04$ for $(V-I) \approx 1.25$ \cite{hol95}). It is not standard according to the average $M^{gc}_{peakI} \approx -8.5$, calculated from the average $M^{gc}_{peakV} \approx -7.5$ \cite{fer00b} and standard globular cluster colors \cite{geb00}.
This initial value was iterated in the process of the SBF distance modulus calculation. 
For the background galaxies,  a power-law luminosity function was used:
\begin{equation}
  N_{bg}(m) = N_{obg} 10^{\gamma m} 
\end{equation}


\begin{table*}[!htf]

\begin{flushleft}
\begin{tabular} {|c|c|c|c|c|c|c|c|c|c|c|c|c|} \hline 

Annulus (arcsec) & P$_0$ & $\sigma_{P_0}$  &$P_{es}$&$(V-I)_o$ &$\overline M_I$&$\overline I_{o,k} $& $\sigma_{\overline I_{o,k}}$&$\overline I_{o,k} - \overline M_I$&$\sigma_{(\overline I_{o,k} - \overline M_I)}$\\ \hline

34 - 40&0.0068 & 0.0003&0.0056 &1.21&-1.45&32.91&0.03 &34.36&0.16 \\ \hline
40 - 46&0.0066 &  0.0006&0.0048 &1.22&-1.45&32.44&0.06 &33.89&0.17\\ \hline
46 - 52&0.0088 & 0.0004&0.0050 &1.20&-1.51&31.42& 0.11  &33.13&0.19\\ \hline
52 - 58&0.0089 &  0.0003&0.0038 &1.18&-1.61&31.10& 0.06 &32.91&0.17 \\ \hline
58 - 64&0.0083 &0.0002& 0.0040&1.16&-1.69&31.30 &0.06 &33.19&0.17 \\ \hline
34 - 64&0.0085&0.0003&0.0054&1.19&-1.54&31.63&0.07&33.17&0.17 \\ \hline
 \hline
 Mean  & -- &-- & --&--&--&--&--&33.44&0.17 \\ \hline

\end{tabular}

\end{flushleft}
\caption{SBF measurements for various annuli of IC 4296} \label{tab-4296annu}
\end{table*}


\noindent with $\gamma = 0.27$  \cite{sma95}.
We kept as fixed parameters in the fit $\sigma$,  $\gamma$, and  $m^{gc}_{peak}$, iterating on the galaxy distance, while
$N_{ogc}$ and $N_{obg}$ were estimated by fitting  the composite luminosity function to the external sources extracted from the image in the range 34$\arcsec$ to 64$\arcsec$.  Identified foreground stars were not included in the fit.
As pointed out by Blakeslee et al. (1999a), the exact details of the adopted luminosity function for the background galaxies and the globular cluster systems have little effect on the final measurement of $m_I$.
From the estimated  $N_{ogc}$ and $N_{obg}$ per pixel, $P_{es}$ was calculated as the sum of \cite{bla95}:
\begin{eqnarray}
\sigma^2_{gc}=&\frac{1}{2} N_{ogc} 10^{0.8[m_1-m^{gc}_{peak}+0.4\sigma^2ln(10)]}\\ \nonumber
& erfc[\frac{m_{cut}-m^{gc}_{peak}+0.8\sigma^2ln(10)}{\sqrt{2}\sigma}]
\end{eqnarray}
\noindent and
\begin{equation}
\sigma^2_{bg}=\frac{N_{obg}}{(0.8-\gamma) ln(10)}10^{0.8(m_1-m_{cut})+\gamma(m_{cut})}.
\end{equation}
\noindent  $m_{1}$ is the zero magnitude which corresponds to a flux of 1 ADUs$^{-1}$.
A $m_{cut} = 24.7$  was used. The corrections were integrated over the luminosity function in each annulus, as in Sodemann \& Thomsen (1995).
To calculate the completeness function, simulated point source images were added to the original, galaxy subtracted image.

\noindent The SBF amplitude is given by:
\begin{equation}
\overline m_I =  -2.5 log(P_0-P_{es}) + m_{1} -\epsilon_{ext} sec(z) \label{eq:mag}
\end{equation}
where $\epsilon_{ext}$ is the extinction coefficient, and $z$ the airmass for the observations.
The color term correction was negligible. We measured $\epsilon_{ext}=0.05 $ and $m_{1}=25.62 \pm 0.04$. The effective airmass was $sec(z)$=1.22.

\subsection[Results on IC 4296]{Results on IC 4296}

The power spectrum fitting of IC 4296 is shown in Fig.~\ref{fig-4296spe}. The results for each annulus are listed in Table~\ref{tab-4296annu}. The errors on SBF magnitudes for each annulus are the standard deviations among different wavelength cuts. The errors due to external source residual contribution subtraction were added in quadrature to the fitting errors. The SBF magnitudes we calculated from Eq.~\ref{eq:mag}, and the errors due to the zero point magnitude and the $\epsilon_{ext}$ were then added in quadrature.

\noindent The SBF magnitudes $I_{o,k}$  were then corrected for galactic absorption assuming $E(B-V) = 0.06$, $A_I=1.940 \ E(B-V)$, and $A_V=3.315 \ E(B-V)$, from Schlegel et al. (1998), and a $k$--correction $k_I\approx 7 \ z$ was applied (Tonry et al. 1997; Liu et al. 2000), $\overline I_{o,k}= \overline m_I - A_I - k_I $. The total $A_I + k_I$ was equal to 0.20 mag.

\noindent The galaxy color in each annulus is shown in Table~\ref{tab-4296annu}. The error on each measure was   $0.03$ mag. The colors that are shown in the table have been corrected for extinction. The adopted $A_I - A_V$ correction amounts to 0.0825 mag.

\noindent From the Tonry et al. (2000) calibration $\overline M_I = (-1.74 \pm 0.08) + (4.5 \pm 0.25) \ [(V-I)_o -1.15] $, we derive the $\overline M_I$ shown in  Table~\ref{tab-4296annu}. The error on each value of $\overline M_I$ is $\sigma_{\overline M_I} = 0.16$.

\noindent In each annulus we derive an estimated distance modulus $\overline I_{o,k} - \overline M_I$. A mean distance modulus was determined as the mean of the values in the considered annuli. The error is given by the standard deviation of those values, around the mean, divided by the square root of the number of considered values. 

\noindent The final distance modulus is $(\overline I_{o,k} - \overline M_I) = 33.44 \pm 0.17$ and the galaxy distance $49 \pm 4$ Mpc. Lauer et al. (1998) have obtained  $(\overline I_{F814} - \overline M_{F814}) = 33.47 \pm 0.13$ with a derived distance of 49 $\pm$ 3 Mpc. The two results are compatible.

\noindent The percentage error that we have on the galaxy distance is $\approx 8\%$. From the theoretical prediction given in Mei et al. (2000), we predict an error percentage of $\approx 10\%$, for a seeing $\approx 0.7\arcsec$/pixel, sky brightness $m^I_{sky}= 19 / $mag/arcsec$^2$, and detection complete at one an half magnitudes below the peak of the globular cluster luminosity function.
The two results are consistent.

\section{Summary and Conclusions}

\noindent  I--band SBF have been observed in IC 4296 with the VLT. A distance modulus of $(\overline I_{o,k} - \overline M_I) = 33.44 \pm 0.17$ was measured, from which a galaxy distance $49 \pm 4$ Mpc is derived. 

\noindent This result confirm the potential of 8--m class telescopes in these kind of measurements and suggest that future SBF observations from the ground can reach the same distance that until now were only reachable by HST.

\begin{acknowledgements}
S. Mei acknowledges support from the European Southern Observatory Studentship programme and Director General's Discretionary Fund. We thank our referee, Laura Ferrarese for her useful comments for the improvement of the paper.

\end{acknowledgements}

\end{document}